\documentclass[aps,floatfix,prd,showpacs]{revtex4}
\usepackage{graphicx}
\usepackage{dcolumn}
\usepackage{bm}

\begin{document}

\def\be{\begin{equation}}
\def\ee{\end{equation}}
\def\dq{\frac{d^4q}{(2\pi)^4}\,}
\def\dqE{\frac{d^4q_E}{(2\pi)^4}\,}

\title{Spin Zero Glueballs in the Bethe-Salpeter Formalism}
\author{Joseph Meyers and Eric S. Swanson}
\affiliation{
Department of Physics and Astronomy, 
University of Pittsburgh, 
Pittsburgh, PA 15260, 
USA.}

\date{\today}

\begin{abstract}
The Schwinger-Dyson Bethe-Salpeter approach to the bound state problem is applied to the spin zero glueball spectrum.
Although a moderately successful description of the two point functions and the glueball spectrum can be obtained, further work is required to reduce sensitivity to model truncations.
\end{abstract}
\pacs{11.10.St,12.38.Lg,12.39.Mk}

\maketitle

\section{Introduction}

The Bethe-Salpeter (BS) formalism is a venerable method for obtaining information on bound state systems in a covariant framework\cite{BS}. It is, however,  technically challenging and most applications to hadronic physics have been made in simple models\cite{BSreview} or in two-body systems\cite{BSmesons}. Only recently has the rather formidable extension to nucleons been undertaken\cite{BSnucleon}.

Because the Bethe-Salpeter formalism is covariant, it is especially useful for describing dynamical quantities such as form factors and distribution functions. It is also a many-body approach and thus can incorporate chiral symmetry breaking, which is crucial to obtaining reliable predictions in the light hadron spectrum\cite{BSmesons}.

Unfortunately, these benefits come at a cost. 
The Bethe-Salpeter formalism hinges on the assumed form of a two-body irreducible scattering kernel, and this form must be obtained with some truncation -- a truncation that must be made with no apparent small parameter in sight. 
This theoretical issue can become acute because truncated kernels often lead to bizarre predictions such as `ghost states'\cite{BSreview}, nonsensical spectra, or unphysical parameter-dependence of observables. Even this undesirable situation can be a distant goal. Often one simply postulates a kernel and tests the accuracy of its predictions. Of course this removes all putative connections to Quantum Chromodynamics (QCD) and the Bethe-Salpeter formalism devolves to a (quite sophisticated) quark model.

Here we explore the extension of the Bethe-Salpeter approach to the gluonic sector of the strong interactions -- namely the glueball spectrum. To our knowledge this is the first attempt at computing glueball masses with the Bethe-Salpeter formalism. It is also a rare example of a hadronic computation that uses QCD to build the interaction kernel\cite{LatModel}.
Thus we seek to leave the realm of pure model-building and enter one in which field-theoretic truncations become the relevant issue. We will, for example, assume that dressed versions of the lowest order kernel dominates the interaction. Thus the ingredients necessary for this investigation are gluon, ghost, and quark propagators and the three-gluon, four-gluon, quark-gluon, and ghost-gluon vertices. It is clear that the problem is formidable, even at this simple level of truncation.

Ensuring the fidelity of the truncations employed is not straightforward.
We are therefore fortunate that a variety of lattice gauge computations of two-point functions have been performed recently. Our strategy will be to model vertices, employ these vertices in Schwinger-Dyson equations for two-point functions, verify the solutions against lattice computations, and then use the resulting suite of $n$-point functions in the glueball Bethe-Salpeter equation. Unfortunately, this is difficult to implement in practice because the gluon two-point function is notoriously truncation-dependent\cite{vSHA,PW}. We therefore choose to use the lattice ghost and gluon propagators directly in the Bethe-Salpeter equation. 

We start by defining the various propagators and vertices used in this work. The two-body Bethe-Salpeter equation for gluonic states is then developed and applied to even charge conjugation scalar glueballs. Model parameters are tuned against lattice gauge theory results and finally, we test the fidelity of the computation by comparing model computations of the relevant two-point functions to lattice results. Although a reasonably good spectrum is obtained, the comparison to two-point functions indicates that substantial truncation error occurs and further work is required to obtain robust QCD-based predictions of hadronic properties in the Bethe-Salpeter formalism.

\section{Propagators and Vertices}

Quark, gluon, and ghost propagators are defined in terms of dressing functions as 
follows:

\begin{eqnarray}
S(k) &\equiv& \frac{i}{A \rlap{/}k - B} \\
D_{\mu\nu}(k) &\equiv& -i\left( g_{\mu\nu} - \frac{k_\mu k_\nu}{k^2}\right)\cdot G(k) - i \xi \frac{k^\mu k^\nu}{k^4+i\epsilon} \\
H^{ab}(k) &\equiv& i \delta^{ab} \, H(k) \equiv i \delta^{ab} \, \frac{h(k)}{k^2}. 
\label{props}
\end{eqnarray}

The scattering kernel that underpins the Bethe-Salpeter formalism can be formally written in terms of Schwinger-Dyson equations. Of course these equations must be truncated at some point. It is traditional to organise the equations in terms $n$-point functions and truncate at the two- or three-point level. We will follow tradition here and thus construct the two particle irreducible four-point function via dressed tree-level interactions. Details will be provided in the next section -- here we focus on the dressed vertices.

\subsection{Model Vertices}

In the case of abelian gauge theory, the Ward-Takahaski identity can be used to completely constrain the longitudinal dependence of the full fermion-gauge boson vertex. The result is given in terms of the fermion dressing functions, $A$ and $B$, and is called the Ball-Chiu vertex\cite{BC}. The situation is more complicated in the nonabelian case because the ghost dressing function and the quark-ghost scattering kernel, $K_g$ appear in the Slavnov-Taylor identity:

\be
p_3^\mu \Gamma^a_\mu(p_1,p_2,p_3) = i g_0 T^a h(p_3)[ S^{-1}(-p_1) K_g(p_1,p_2,p_3) - \bar{K_g}(p_2,p_1,p_3) S^{-1}(p_2)].
\ee

The resulting vertex has a rich structure that was studied by Aguilar and Papavasilliou\cite{AP}. They found that an adequate approximation to the full longitudinal structure of the vertex can be obtained by neglecting the quark-gluon scattering kernel and multiplying the Ball-Chiu vertex by the ghost dressing function. We make the further simplification of retaining only the `central' term (proportion to a gamma matrix) of the Ball-Chiu vertex:
\be
\Gamma_\mu^a(p_1,p_2,p_3) = - i g_0 T^a \,\gamma_\mu\, h^n(p_1-p_2)\, \frac{1}{2}[A(p_1)+A(p_2)].
\label{qg-vertex}
\ee

A well-known shortcoming of the Ball-Chiu approach is that the transverse part of the vertex function remains unspecified. This has been partly remedied in the Abelian case by requiring multiplicative renormalisability in the asymptotic regime\cite{CP}. No such Ansatz exists for the nonabelian case and we choose to neglect this contribution henceforth. However, Aguilar and Papavasilliou note that the correct perturbative asymptotic behaviour for the quark propagator is not obtained with this approximation -- presumably due to the missing transverse components of the vertex. They then observe that supplying an extra power of the ghost propagator in Eq. \ref{qg-vertex} ($n=2$) allows recovery of the expected asymptotic behaviour. We therefore choose to allow for varying values of $n$ in the following.

The ghost-gluon vertex has been considered by von Smekal {\it et al.}\cite{vSHA} who 
neglect the ghost-ghost scattering contribution to the vertex and retain only the reducible part of the four-ghost scattering matrix element. The result is an Ansatz of the form

\be
{\Gamma_\mu^{abc}}(q,p) =  g_0 f^{abc}  \left[q_\mu \frac{h(k)}{h(q)} +  p_\mu \left( \frac{h(k)}{h(p)} -1\right)\right]
\label{eqn-H}
\ee
where $q$ is the incoming ghost momentum, $p$ is the outgoing ghost momentum, and $k = p-q$ is the incoming gluon momentum.

The authors of Ref. \cite{vSHA} also examine the three-gluon vertex. They employ the 
ghost scattering kernel implicit in Eq. \ref{eqn-H} to obtain an Ansatz as follows:

\be
\Gamma(123) = g_0 f^{123}\left[ A_+(123) g_{12} (p_1-p_2)_3 + A_-(123)g_{12}(p_1+p_2)_3 + 2 \frac{A_-(123)}{p_1^2-p_2^2}\, [g_{12} p_1\cdot p_2 - (p_1)_2 (p_2)_1] (p_1-p_2)_3] \ {\rm et\ cyc.} \right]
\ee
where
\be
A_\pm(123) = \frac{h(p_3)}{2} \left( \frac{h(p_2)}{h(p_1)G(p_1) p_1^2} \pm \frac{h(p_1)}{h(p_2)G(p_2) p_2^2}\right).
\label{3g-vertex}
\ee
This form ignores the unconstrained transverse components, but these are known to be suppressed\cite{BP}.
Since one expects $A_- \ll A_+$ we simplify the vertex further be neglecting the terms proportional to $A_-$.

An alternative model is advocated by Pennington and Wilson\cite{PW} who employ an approximate solution to the Slavnov-Taylor identity with a bare ghost-gluon scattering kernel:

\be
\tilde A_+(123) = \frac{h(p_3)}{2} \left( \frac{1}{G(p_1)p_1^2} + \frac{1}{G(p_2)p_2^2}\right).
\label{eq-PW}
\ee

Vertex models for the four-gluon vertex have been rarely considered\cite{V4g}; we simply use the perturbative vertex in the following.

\subsection{Gap Equations}

The propagator dressing functions of Eqs. \ref{props}
are obtained by solving the two-point Schwinger-Dyson equations. These require renormalisation which we implement with $Z$-factors as defined in the following Lagrangian:

\begin{eqnarray}
{\cal L}_0 &=& -\frac{1}{4}Z_A (\partial_\mu A_\nu^a - \partial_\nu A_\mu^a)^2  + i Z_F \bar \psi \rlap{/}\partial \psi 
- Z_F m_0 \bar\psi \psi - Z_c {\bar c}^a \partial^2 c^a \nonumber \\
&&  + g_0 Z_F Z_A^{1/2} \bar \psi \rlap{/}A \psi - g_0 Z_A^{3/2} f^{abc}(\partial_\mu A_\nu^a) A_\mu^b A_\nu^c - g_0^2 Z_A^2 (f^{eab}A_\mu^a A_\nu^b)(f^{ecd} A_\mu^c A_\nu^d) \nonumber \\
&& -g_0 Z_c Z_A^{1/2} f^{abc}{\bar c}^a \partial_\mu A_\mu^b c^c - \frac{1}{2\xi} (\partial_\mu A_\mu^a)^2.
\end{eqnarray}

\begin{figure}[ht]
\includegraphics[width=4cm,angle=0]{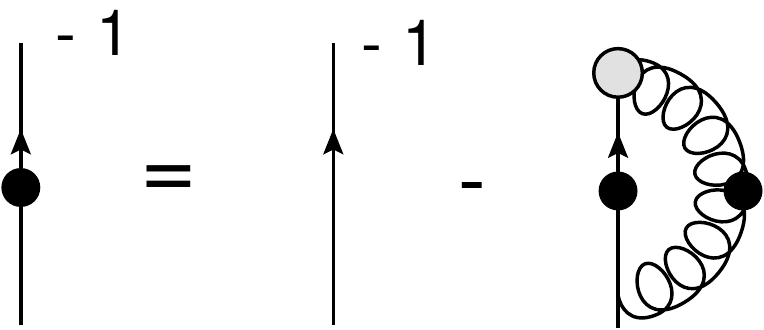}
\qquad\qquad
\qquad\qquad
\includegraphics[width=4cm,angle=0]{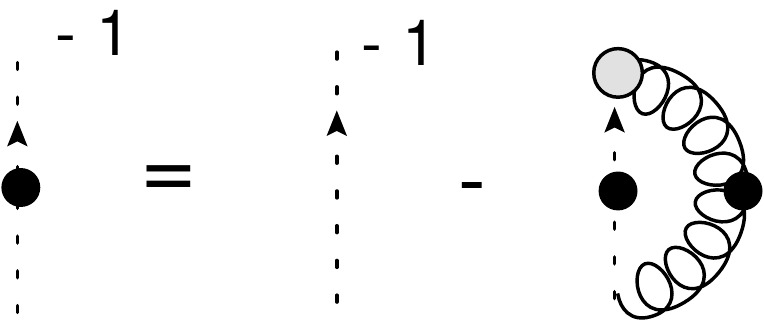}
\caption{Quark and Ghost Gap Equations.}
\label{fig-gap}
\end{figure}

The exact quark and ghost gap equations are shown in Fig. \ref{fig-gap}. As discussed above, these are truncated by employing vertex models.  Projecting the full quark propagator onto its two components, working in Landau gauge,  and performing the Wick rotation to Euclidean space yields the coupled gap equations:

\be
B(p_E^2) = Z_F m_0 + 3 g_0^2 Z_F^2 Z_A C_F \int \dqE W\, \frac{B(q_E^2) G_E(Q_E)}{A^2 q_E^2 + B^2},
\label{eq-B-gap}
\ee

\be
A(p_E^2) = Z_F - g_0^2 Z_F^2 Z_A C_F \int \dqE W\,\frac{A(q_E^2) G_E(Q_E)}{A^2q_E^2 + B^2}\,\left(\frac{2 q_E^2(1-x^2)}{Q_E^2} - 3 \frac{q_E x }{p_E}\right),
\label{eq-A-gap}
\ee
and

\be
\frac{1}{h(p_E^2)} = Z_c -  g_0^2 Z_c^2 Z_A C_A\, \int \frac{q_E^3dq_E}{4\pi^3}\, \sqrt{1-x^2} dx\, V  G_E(Q_E) h(q_E^2) \frac{1-x^2}{Q_E^2}.
\label{eq-h-gap}
\ee
Here $Q=p-q$ and the Euclidean nature of these expressions has been emphasised with the subscript, which we subsequently suppress. By convention $G_E(Q_E) = - G(Q \to Q_E)$, and $B(p_E) = B(p\to p_E)$.
Factors due to vertex models are given as

\be
V = \frac{h(Q)}{h(q)} + \frac{h(Q)}{h(p)} - 1
\label{eq-V}
\ee
and 
\be
W = h^n(Q) \frac{1}{2} (A(p) + A(q)).
\ee
Rainbow-ladder results are recovered when $h\to 1$ and $A\to 1$ in these expressions.

\begin{figure}[ht]
\includegraphics[width=10cm,angle=0]{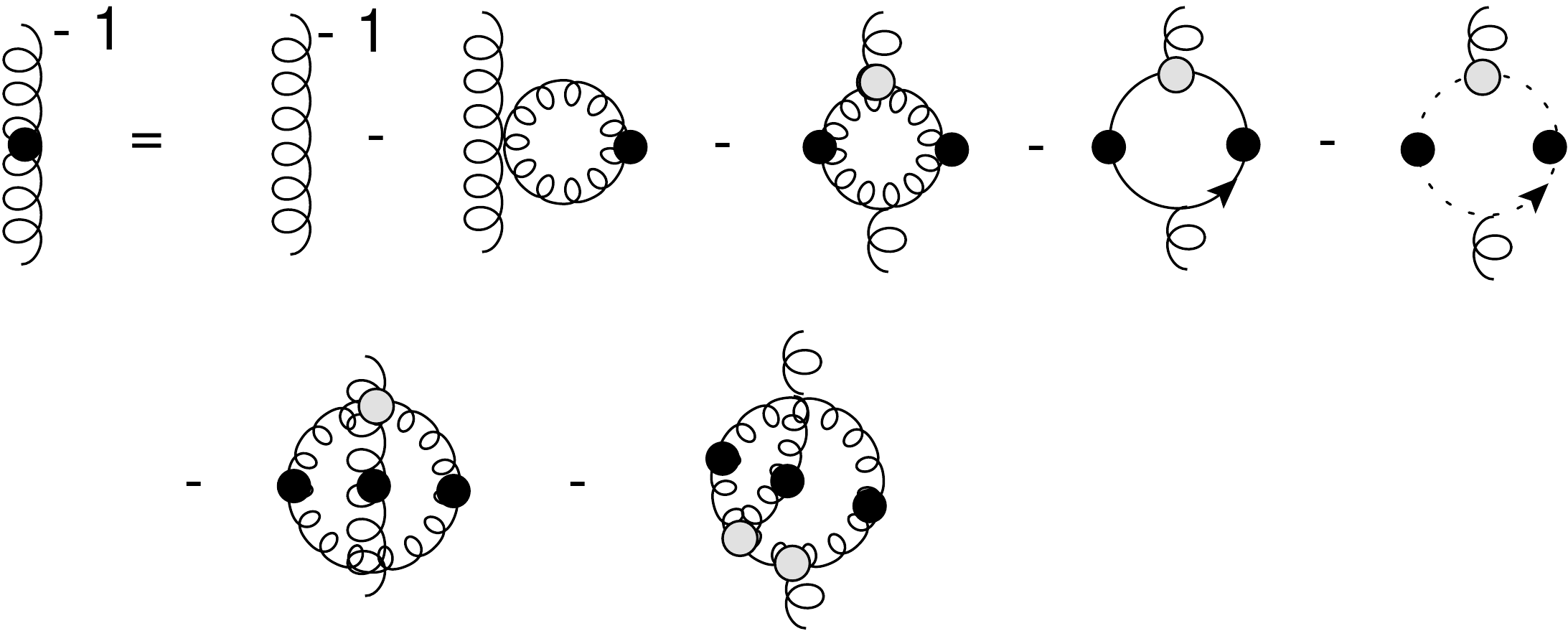}
\caption{Gluon Propagator Schwinger-Dyson Equation. Solid circles represent full propagators. The open circles represent the full vertex.}
\label{fig-gluon-gap}
\end{figure}

The gluon gap equation, shown in Fig. \ref{fig-gluon-gap}, has a rich structure that makes obtaining definitive numerical solutions notoriously difficult\cite{PW,AP,vSHA}. We therefore adopt the simple expedient of employing a fit to the lattice gauge results. It is important to note that this is not just a simple matter of finding a parameterization that mimics the data in a reasonable way because solving the glueball Bethe-Salpeter equation requires knowledge of the gluon propagator in the Minkowski regime. This will be discussed in greater detail below.

We chose to renormalise the quark and ghost gap equations by setting
\be
g_0 Z_F Z_A^{1/2} =g.
\label{eq-g-renorm}
\ee
However, in Landau gauge
\be
g_0 Z_c Z_A^{1/2} = g
\label{eq-g-ghost}
\ee
holds to all orders\cite{JCT}. Thus gauge covariance implies that $Z_c(\mu,\Lambda) = Z_F(\mu,\Lambda)$, where $\Lambda$ is the scale associated with the regularisation scheme. Truncation and model error break gauge covariance and thus the equality of the ghost and quark field strength renormalisation factors cannot be expected. This relationship will be used as a measure of the fidelity of the model below.

We remark that the form of Eqs. \ref{eq-B-gap} -- \ref{eq-h-gap} follows tradition in that vertex renormalisation factors are made explicit. However, this is in fact not necessary, or even desirable: once a vertex model has been adopted it can be regarded as renormalised (just like the quark dressing functions) and hence the renormalisation factors can simply be absorbed into the vertex factors $V$ and $W$. Our prescription is equivalent to making this choice.

Once Eqs. \ref{eq-g-renorm} and \ref{eq-g-ghost} are in place, renormalisation can be completed by performing a single subtraction at a scale $\mu$ to determine $Z_c(\mu,\Lambda)$ and $Z_F(\mu,\Lambda)$. We remark that the subtracted gap equations are 
often renormalised at large Euclidean momentum where it is assumed that the perturbative form of the quark propagator is recovered. In general this is not true, and we choose to retain the constants $B(\mu)$, $A(\mu)$, and $h(\mu)$ as possible input parameters of the theory.

%
%
%

\subsection{Numerical Results and Comparison to the Lattice}

\begin{figure}[ht]
\includegraphics[width=10cm,angle=0]{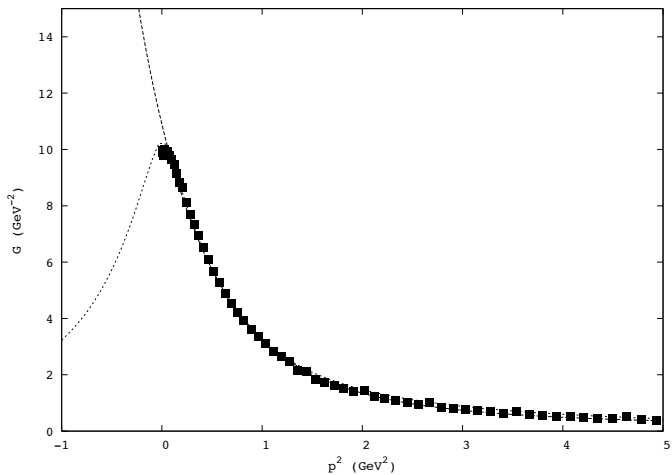}
\caption{Lattice results\protect\cite{lattice-gluon} for the Landau gauge gluon propagator and model fits. Error bars are smaller than the points. Dashed line Eq. \ref{eq-B-fit-1}; dotted line Eq. \ref{eq-B-fit-2}.}
\label{fig-gluon}
\end{figure}

The primary ingredient in this investigation is the gluon dressing function, $G$. The unrenormalised propagator  has been computed in Landau gauge in pure $SU(3)$ gauge theory on large lattices\cite{lattice-gluon}, with results shown in Fig. \ref{fig-gluon}.

A good fit to the lattice data is obtained with

\be
G_{\rm lat}(p^2) = 1.409\, \frac{1 - \exp(-2.00\,(p^2 - 1.144))}{p^2 - 1.144}\ {\rm GeV}^{-2}.
\label{eq-B-fit-1}
\ee
The figure also displays an alternative fit of the form

\be
G_{\rm lat}(p^2) = \frac{4.70}{(p^2)^{1.44} + 0.454} \ {\rm GeV}^{-2},
\label{eq-B-fit-2}
\ee
which was specifically chosen to be similar to Eq. \ref{eq-B-fit-1} in the `Euclidean region' ($p_E^2 > 0$) but to differ dramatically in the `Minkowski region' ($p_E^2 < 0$).
Notice that the expected asymptotic behaviour $G \to 1/p^2 +\ {\rm logs}$ is not incorporated in the fits. This should not affect glueball properties or the long range structure of the propagators since these are dominated by the infrared regime. Although the lattice data carry units, the propagator has not been normalised\cite{lattice-gluon}; we therefore must allow for a multiplicative renormalisation factor in subsequent computations involving the gluon propagator. Finally, as indicated in the figure, these fits yield essentially identical results in the Euclidean region, but differs crucially in the Minkowski region, $p^2<0$. The importance of this will be discussed in section \ref{sect-sg}.

\begin{figure}[ht]
\includegraphics[width=10cm,angle=0]{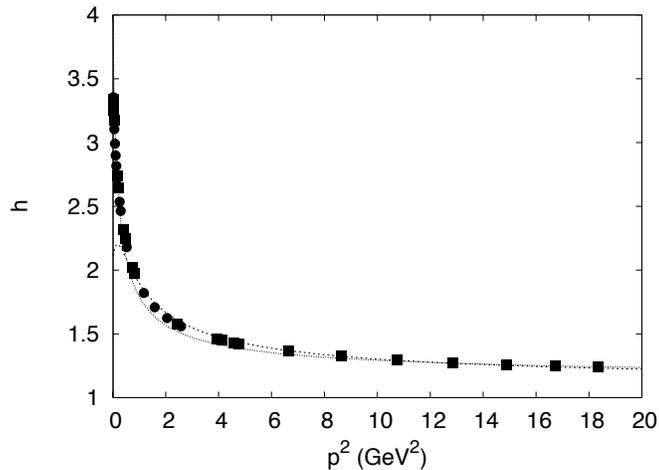}
\caption{The renormalised ghost dressing function and lattice results\protect\cite{lattice-gluon}. Error bars are smaller than the data points. Dashed line: bare vertex model; dotted line: full vertex model.}
\label{fig-ghost}
\end{figure}

The ghost propagator is computed next since it may be obtained self-consistently with an Ansatz gluon propagator.
Even though the ghost lattice data are unrenormalised, we chose to renormalise with respect to them so that a reasonable comparison can be made. In this case we set $\mu = 3.57$ GeV and $h(\mu) = 1.27$. Lattice data and numerical solutions to Eq. \ref{eq-h-gap} are displayed in Fig. \ref{fig-ghost}. We chose to scale the gluon propagator so that $G_R(\mu_g) = 1/\mu_g^2$, thus

\be
G_R(p;\mu_g) = \frac{G_{\rm lat}(p)}{G_{\rm lat}(\mu_g) \mu_g^2}.
\ee
The coupling $g^2(\mu_g)$ is then fit to the lattice result to obtain the final numerical prediction for the ghost propagator. The result for the case of the bare ghost-gluon vertex is shown as a dashed line in the figure. In this case the gluonic renormalisation scale was set to 1 GeV and the fit coupling was $g(1) = 2.4$. Notice that since the coupling and gluon propagator occur together in the gap equation, the scale dependence of the coupling is simply given by

\be
g^2(\mu_g) = g^2(1) \frac{G_{\rm lat}(\mu_g) \mu_g^2}{G_{\rm lat}(1)}
\ee
where $\mu_g$ is measured in GeV.

The quality of the fit is quite good over the entire range of the propagator, although it does worsen near the origin. An alternative fit using the vertex model of Eq. \ref{eqn-H} (or Eq. \ref{eq-V}) is shown as a dotted line in the figure. In this case the propagator saturates in the infrared and does not produce the sharp peak seen in the lattice data. We therefore prefer to use the bare ghost-gluon vertex in the following.

Finally, we consider the quark dressing functions.  We compare to the lattice computation of Ref. \cite{lattice-quark} where the mass function $M=A/B$ and the wavefunction renormalisation function $Z = 1/A$ were computed in quenched and unquenched Landau gauge QCD with 2+1 flavours. In this case the mass function does not depend on the renormalisation parameters and a direct comparison can be made. Alternatively, the authors of Ref. \cite{lattice-quark} chose to renormalise $A$ by setting $Z(p= 3\ {\rm GeV}) = 1$, which we also employ.

\begin{figure}[ht]
\includegraphics[width=10cm,angle=0]{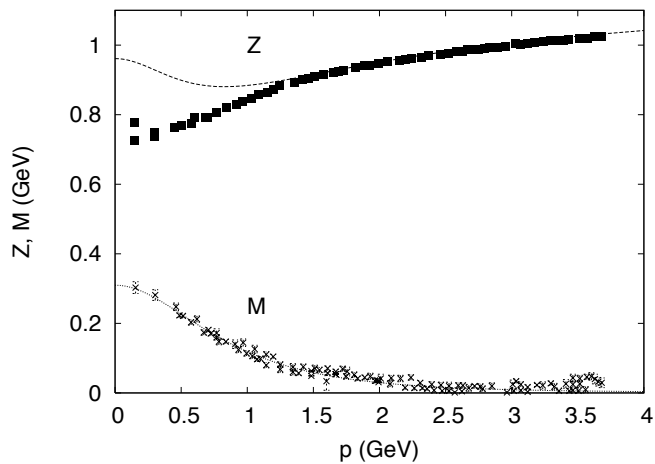}
\caption{The quark wavefunction renormalisation, $Z$, and quark running mass $B$, along with lattice results\protect\cite{lattice-quark}.} 
\label{fig-quark}
\end{figure}

Results of the computation are shown in Fig. \ref{fig-quark}. The lattice data are unquenched results in the chiral limit. The fit coupling is $g(1\ {\rm GeV}) = 4.8$ and the modified central Ball-Chiu vertex with $n=2$ has been employed. We have found that fits with $n=1$ give similar results but with larger couplings.
Again, the agreement is satisfactory, although the wavefunction renormalisation deviates from the lattice data by 20\% in the infrared region.

\section{Glueball Bethe-Salpeter Equation}

\subsection{Notation and General Expressions}

The Bethe-Salpeter equation is traditionally derived by considering  four-point scattering near resonance locations via an equation that can be written  schematically as

\be
\Gamma_4 = \sum_R \Gamma_R \frac{1}{s-m_R^2} \Gamma_R + \ {\rm regular}.
\label{eq-reg}
\ee
The vertex factors are the Bethe-Salpeter amplitudes that describe the bound state resonance in question. In the case of gluon-gluon scattering the  BS amplitude is given by

\be
\Gamma_R^{\mu\nu}(x,y;P) = \langle 0| T A^\mu(x)A^\nu(y)|R(P,\lambda)\rangle
\label{eq-me}
\ee
where the ket labels the glueball in terms of its four-momentum and (possible) polarisation index. The Fourier transform of the BS amplitude is defined as

\be
\Gamma^{\mu\nu}(x,y;P) = {\rm e}^{-iP\cdot \frac{x+y}{2}} \, \int \frac{d^4k}{(2\pi)^4} \, {\rm e}^{-ik\cdot(x-y)} \, \Gamma^{\mu\nu}(k,P)
\ee
where $k$ is the relative four-momentum describing the two-body state. It is convenient to introduce a modified BS amplitude, $\chi$, via

\be
\Gamma_{\mu\nu}(k,P) = D_{\mu\alpha}(k_+) \chi_{\alpha\beta}(k_+,k_-) D_{\beta\nu}(k_-),
\ee
where 
\be
k_\pm = \frac{P}{2} \pm k. 
\ee

Bethe-Salpeter equations are typically simplified by projecting the BS amplitudes onto different Dirac or Lorentz structures. Thus, for example, the pseudoscalar quark amplitude can be expanded as\cite{chls}

\be
\rlap{$\chi$}\,\chi(k,P) = \gamma_5 \{ iE + \rlap{/}k F + \rlap{/}P G + i [\rlap{/}k,\rlap{/}P] H\}.
\label{eq-pion}
\ee
We derive similar expansions by considering transversality and parity constraints on the gluonic Bethe-Salpeter amplitude.

Transversality of the gauge field implies that $k_+^\mu \Gamma_{\mu\nu}(k,P) = k_-^\nu \Gamma_{\mu\nu}(k,P) = 0$. If one considers the gauge $\xi =\infty$ one obtains the transversality conditions

\be
k_+^\mu \chi_{\mu\nu}(k_+,k_-) = k_-^\nu \chi_{\mu\nu}(k_+,k_-) = 0.
\ee

Application of the parity operator to the matrix element in Eq. \ref{eq-me} yields the following relationship
\be
\chi^{\mu\nu}_R(k_+,k_-) = \epsilon(\mu,\nu) \cdot \Pi_R \cdot \chi^{\mu\nu}_R(\tilde k_+,\tilde k_-)
\ee
where $\Pi_R$ is the parity of the resonance $R$,
 $\tilde k = (k_0,-\vec k)$, and $\epsilon$ is a matrix of signs with values $\epsilon(0,0) = \epsilon(i,j) =1 $ and $\epsilon(0,i) = \epsilon(i,0) = -1$.

The general Lorentz structure for a scalar glueball must be
\be
\chi^{\mu\nu}(k_+,k_-) = A_0 g^{\mu\nu} + A_1 k_+^\mu k_+^\nu + A_2 k_-^\mu k_-^\nu + A_3 k_+^\mu k_-^\nu + A_4 k_-^\mu k_+^\nu + A_5 \epsilon^{\mu\nu\alpha\beta} {k_+}_\alpha {k_-}_\beta.
\ee
Applying the transversality and parity constraints then yields the relationships

\be
\chi^{\mu\nu}(k_+,k_-) = \epsilon^{\mu\nu\alpha\beta}k_\alpha P_\beta F(k,P) \quad [0^{-+}]
\label{eq-zeromp}
\ee
for the pseudoscalar glueball
and 
\be
\chi^{\mu\nu}(k_+,k_-) = A(k,P) A^{\mu\nu} + B(k,P) B^{\mu\nu} \quad [0^{++}]
\label{eq-zeropp}
\ee
for the scalar glueball.
Here

\be
A_{\mu\nu} \equiv \frac{k_\mu^{\perp +}  k_\nu^{\perp -}}{k^{\perp +} \cdot k^{\perp -}},
\ee

\be
B_{\mu\nu} \equiv g_{\mu\nu}  -\frac{k_\mu^- k_\nu^+}{k^+\cdot k^-},
\ee

\be
k_{\mu}^{\perp +} = k_\mu^+ - k_\mu^- \frac{(k^+)^2}{k^+\cdot k^-},
\ee
and
\be
k_{\mu}^{\perp -} = k_\mu^- - k_\mu^+ \frac{(k^-)^2}{k^+\cdot k^-}.
\ee

\begin{figure}[ht]
\includegraphics[width=8cm,angle=0]{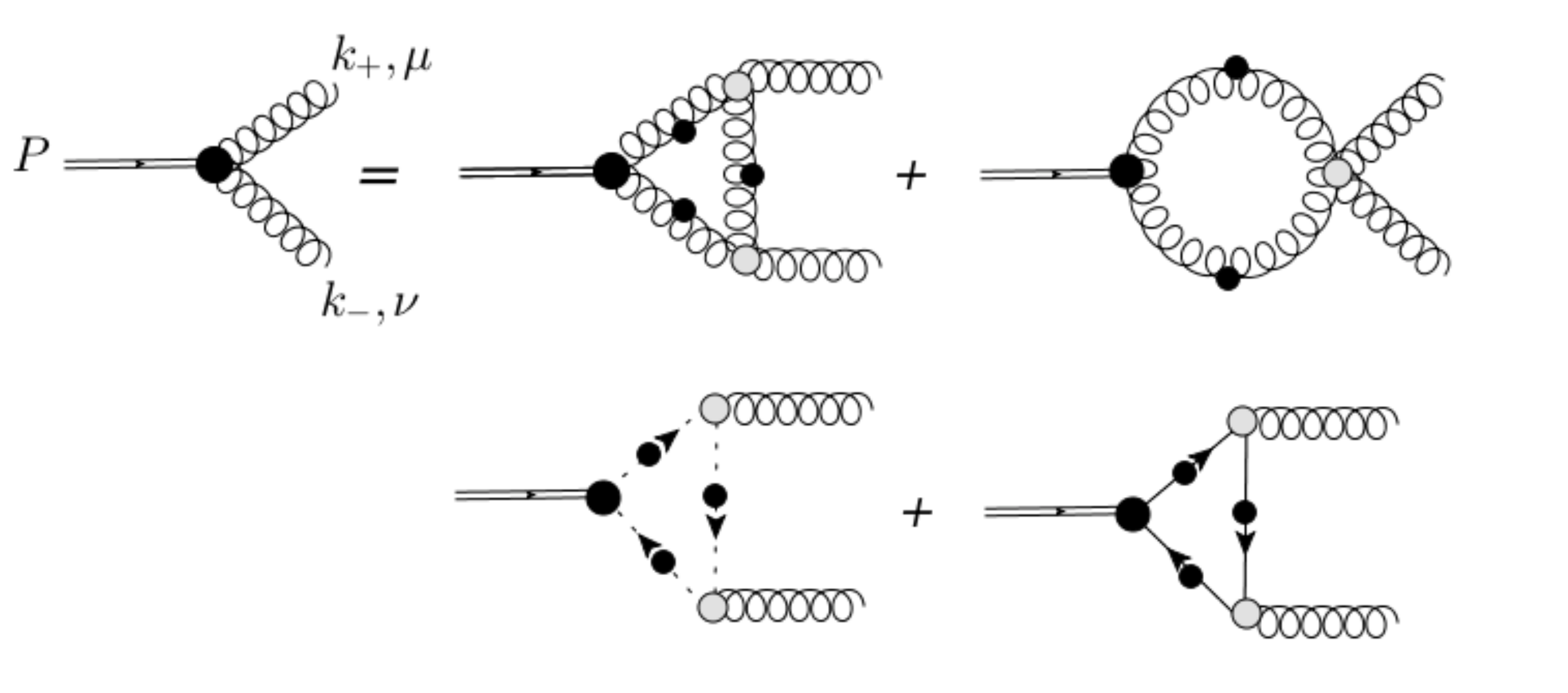}
\caption{Gluonic Bethe-Salpeter Equation. Dots represent full propagators and model vertices. Crossed diagrams are not shown.}
\label{BSE-g}
\end{figure}

\begin{figure}[ht]
\includegraphics[width=8cm,angle=0]{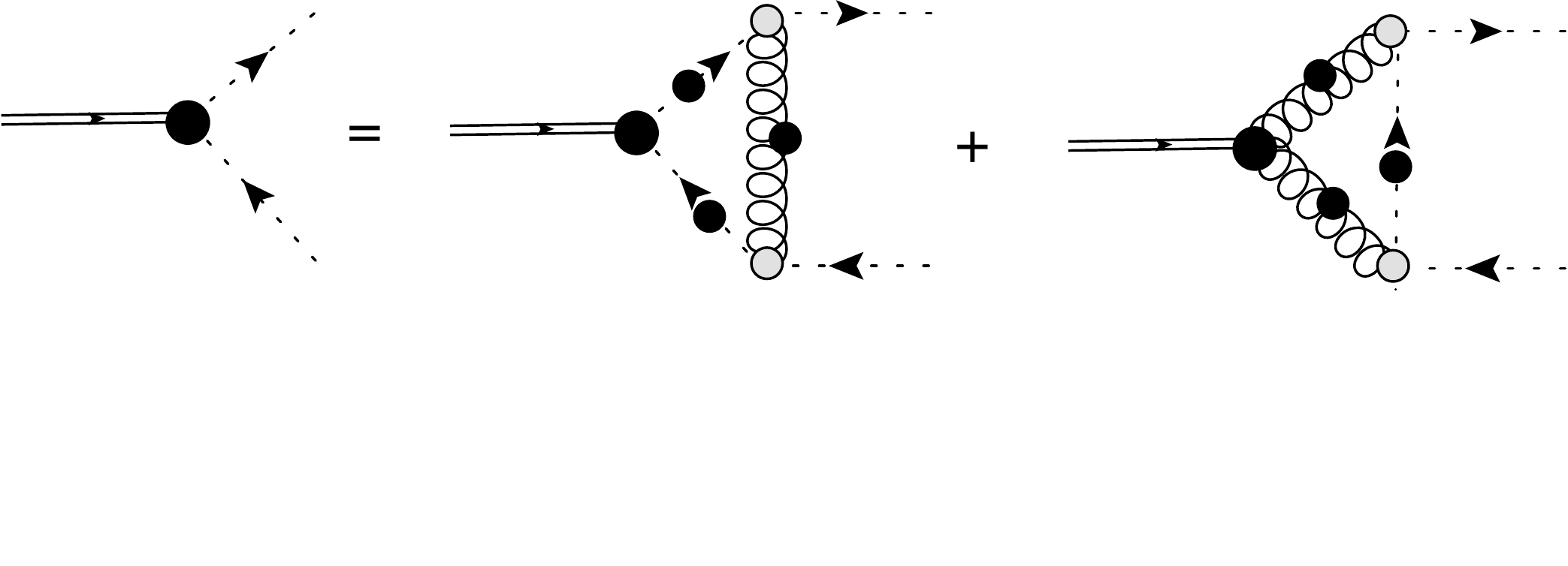}
\caption{Ghost Bethe-Salpeter Equation.}
\label{BSE-gh}
\end{figure}

\begin{figure}[ht]
\includegraphics[width=8cm,angle=0]{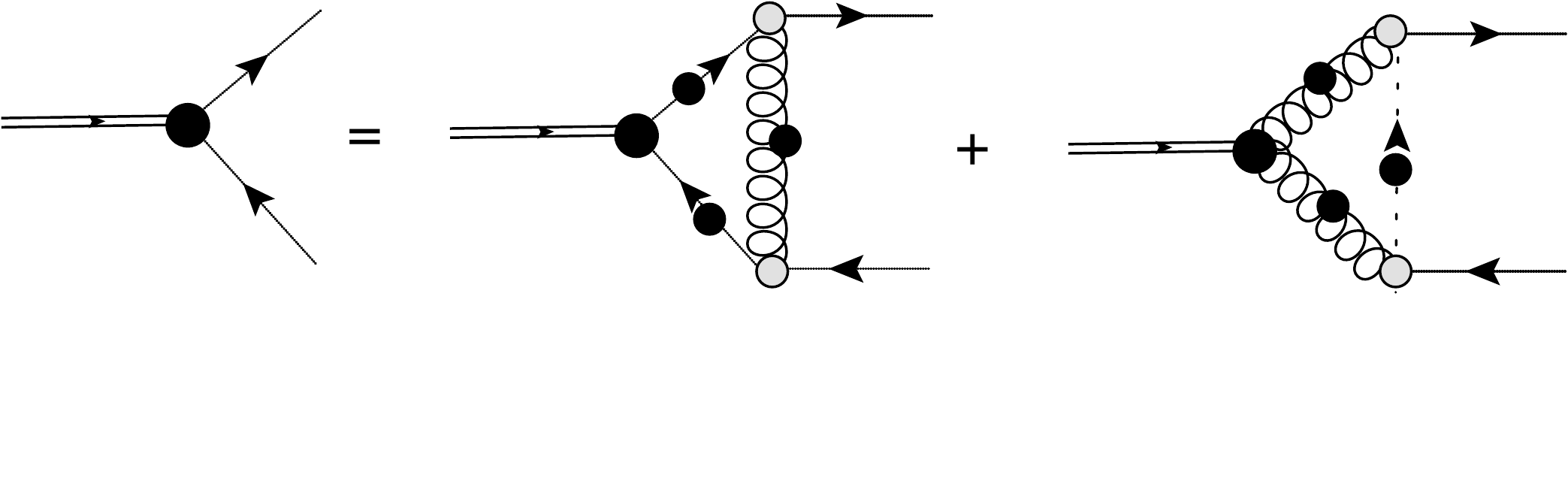}
\caption{Quark Bethe-Salpeter Equation.}
\label{BSE-q}
\end{figure}

\begin{figure}[ht]
\includegraphics[width=8cm,angle=0]{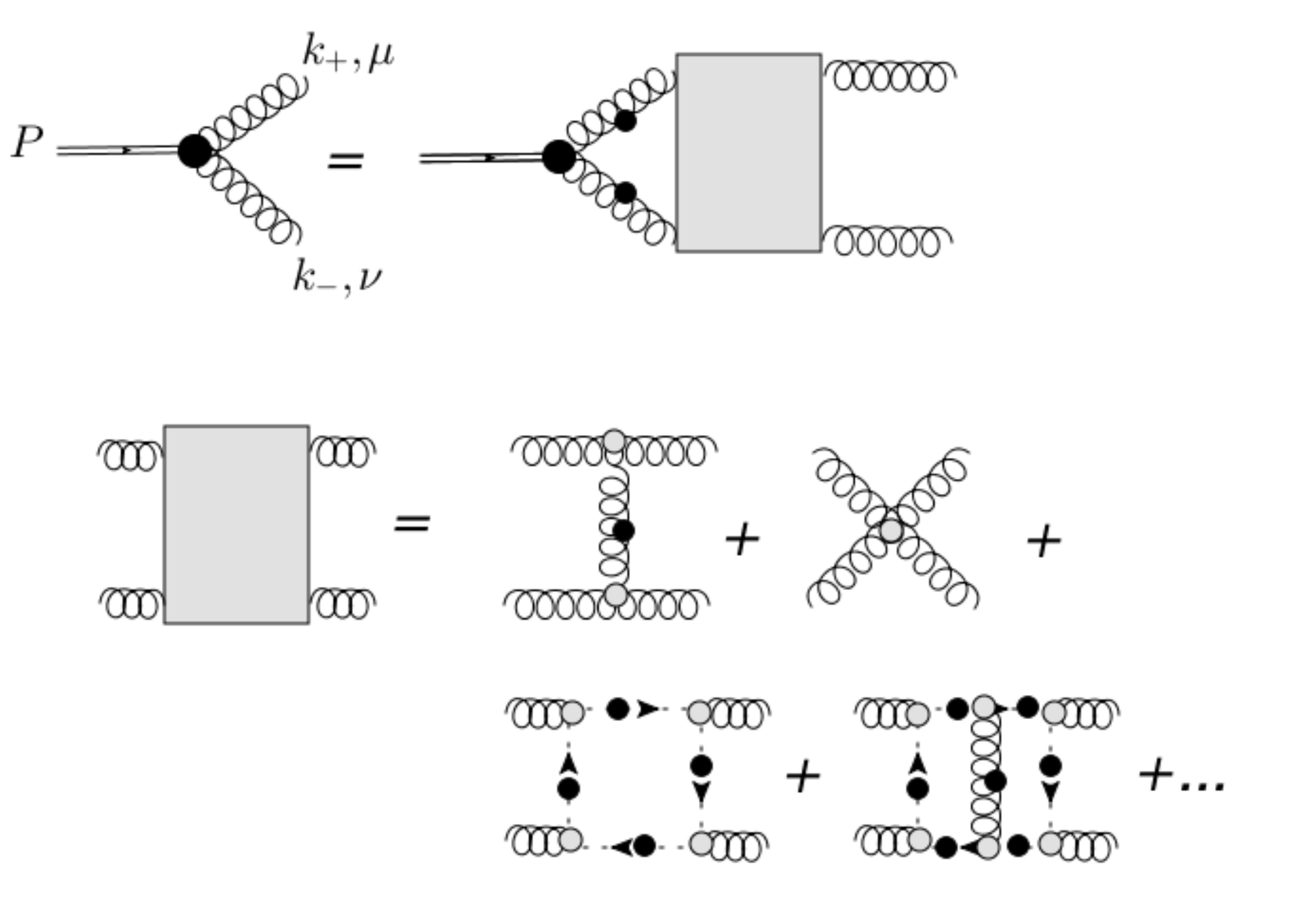}
\caption{Iterated Bethe-Salpeter Equation indicating the BS kernel in the pure gauge case.}
\label{BSE-all}
\end{figure}

The Bethe-Salpeter equation considered here is displayed in Figs. \ref{BSE-g}, \ref{BSE-gh}, and \ref{BSE-q}. Notice that quark and ghost Bethe-Salpeter amplitudes appear in the first equation (Fig. \ref{BSE-g}). This may surprise the reader since no such amplitudes figure in the expansion of the scattering amplitude alluded to above (Eq. \ref{eq-reg}). But the equations for the quark and ghost BS amplitudes can be substituted into the gluon equation to obtain the  expression indicated diagrammatically in Fig. \ref{BSE-all}. This equation follows the usual BS conventions and indicates that the kernel is given by an infinite collection of diagrams that subsumes gluon exchange, the gluonic contact interaction, and dressed quark and ghost box diagrams.  

Although the resulting BS kernel is quite sophisticated it represents an extreme truncation of the diagrams that contribute to the gluonic bound state. For example, one can speculate that diagrams in which many gluons are emitted from the BS amplitude are important at long distances. Nevertheless, we shall press ahead to determine if this relatively simple model can yield a spectrum with any resemblance to lattice data.

In the following the ghost BS amplitude will be denoted $\chi$, while the quark BS amplitude will be denoted $\rlap{$\chi$}\,\chi$. 
Then the coupled Minkowski gluon-ghost-quark system in Landau gauge is given by the following equations.

\begin{eqnarray}
\label{eq-bse-g}
\chi_{\mu\nu}(k_+,k_-) &=& i g^2 N \int\frac{d^4q}{(2\pi)^4} \, \chi_{\alpha\beta}(q_+,q_-) \, {\cal C}^{\alpha\beta}_{..\mu\nu} \, G(q_+)G(q_-) \nonumber \\
&& +  i g^2 N \int\frac{d^4q}{(2\pi)^4} \, \chi_{\alpha\beta}(q_+,q_-) \, {\cal T}^{\alpha\beta}_{..\mu\nu}(q,k;P) \, G(q_+) G(q_-)G(Q) \nonumber \\
&& +  i g^2 N \int\frac{d^4q}{(2\pi)^4} \, \chi(q_+,q_-) \, {\cal G}_{\mu\nu}(q,k;P) \, H(q_+) H(q_-) H(Q) \nonumber \\
&& - \frac{g^2}{2} \int \dq {\rm tr}\left[ \gamma_\mu S(q_+) \rlap{$\chi$}\,\chi(q_+,q_-) 
S(-q_-) \gamma_\nu S(Q)\right]\, h^n(k_+)h^n(k_-) \bar A(Q,q_+)\bar A(Q,q_-)  \nonumber \\
&& + \ {\rm crossed} \\
\label{eq-bse-gh}
\chi(k_+,k_-) &=& i g^2 N \int \frac{d^4q}{(2\pi)^4}\, \chi(q_+,q_-) \, {\cal H}(q,k;P)\, H(q_+) H(q_-) G(Q) \nonumber \\
&& + i g^2 N \int \frac{d^4q}{(2\pi)^4}\, \chi_{\alpha\beta}(q_+,q_-) \, {\cal B}^{\alpha\beta}(q,k;P)\, G(q_+) G(q_-) H(Q)\\
\label{eq-bse-q}
\rlap{$\chi$}\,\chi(k_+,k_-) &=& g^2 C_F \int \dq \gamma_\alpha S(-Q)\gamma_\beta G(q_+) G(q_-) \chi_{\alpha\beta}(q_+,q_-) h^n(q_+)h^n(q_-) \bar A(Q,k_+) \bar A(Q,k_-) \nonumber\\
 && + ig^2 C_F \int \dq \gamma_\mu S(q_+) \rlap{$\chi$}\,\chi(q_+,q_-) S(-q_-) \gamma_\nu G(Q) P_{\mu\nu}(Q)\, h^{2n}(Q)\, \bar A(k_+,q_+)\, \bar A(k_-,q_-).
\end{eqnarray}
where $Q = q-k$ and $\bar A(k,p) = (A(k^2)+A(p^2))/2$.

The tensors in these expressions are as follows. The gluonic contact term is

\be
{\cal C}^{\alpha\beta\mu\nu} = 2 g^{\mu\nu}g^{\alpha\beta} - g^{\alpha\nu}g^{\beta\mu} - g^{\alpha\mu}g^{\beta\nu}.
\ee
The three-gluon interaction with the vertex of Eq. \ref{3g-vertex} is given by

\be
{\cal T}_{\alpha\beta\mu\nu} = P^{\gamma\gamma'}(Q)\, V_{\alpha\mu\gamma}(q_+,k_+)\, V_{\beta\nu\gamma'}(q_-,k_-),
\ee
where the projection tensor is 
\be
P^{\gamma\gamma'}(Q) = g^{\gamma\gamma'} - \frac{Q^\gamma Q^{\gamma'}}{Q^2}
\ee
and the three-gluon vertex tensor is
\begin{eqnarray}
V_{\alpha\mu\gamma}(q_+,k_+) &=& A_+(q_+,k_+,q_+-k_+)g^{\alpha\mu}(q_++k_+)^\gamma + A_+(k_+,q_+-k_+,q_+) g^{\mu\gamma}(q_+-2k_+)^\alpha + \nonumber \\ 
&& +A_+(q_+-k_+,q_+,k_+) g^{\gamma\alpha}(k_+ - 2 q_+)^\mu.
\end{eqnarray}
Here we reference the vertex factor, $A_+$,  of Eq. \ref{3g-vertex} or Eq. \ref{eq-PW}.

The ghost-gluon coupling tensor is given by 
\be
{\cal G}^{\mu\nu} = \left(q_+^\mu \frac{h(k_+)}{h(q_+)} + Q^\mu \left(\frac{h(k_+)}{h(Q)}-1\right)\right) \left( Q^\nu \frac{h(k_-)}{h(Q)} - q_-^\nu \left(\frac{h(k_-)}{h(q_-)}-1\right)\right) .
\ee

Finally, the tensors appearing in the ghost BS amplitude equations are

\be
{\cal H} = P_{\mu\nu}(Q) k_+^\mu k_-^\nu \cdot \left(\frac{h(Q)}{h(k_+)} + \frac{h(Q)}{h(q_+)} - 1\right)\cdot \left( \frac{h(Q)}{h(k_-)} + \frac{h(Q)}{h(q_-)} -1 \right)
\ee
and
\be
{\cal B}^{\alpha\beta} = -k_+^\alpha k_-^\beta \cdot \left( \frac{h(q_+)}{h(Q)} + \frac{h(q_+)}{h(k_+)}-1\right) \cdot \left( \frac{h(q_-)}{h(Q)} + \frac{h(q_-)}{h(k_-)} -1 \right).
\ee

\subsection{Spin Zero Glueballs}
\label{sect-sg}

The spin zero (even charge conjugation) spectrum is obtained by substituting the pseudoscalar and scalar expressions for the gluonic Bethe-Salpeter amplitudes, Eqs. \ref{eq-zeromp} and \ref{eq-zeropp}, into Eqs. \ref{eq-bse-g} -- \ref{eq-bse-q}. Equations for the unknown amplitudes are obtained by projecting where appropriate and solving an eigenvalue problem of the type 
\be
\vec F(k;P) = \lambda(P) \int d^4q\, {\cal K}(k,q;P) \vec F(q;P).
\ee
The eigenvectors and eigenvalues depend parametrically on the glueball momentum $P_E^2$, which is varied to obtain the glueball mass at $\lambda(P_E^2 = -M^2) = 1$. 

Recall that the BS formalism is developed with the Euclidean metric, which means that the mass condition just specified can only be obtained with imaginary Euclidean glueball momenta. 
This presents a potentially daunting obstacle since variables such as $q_\pm$ and $k_\pm$ become complex and hence dressing functions must be known over (a portion of) the complex plane. Note that this problem does not arise in the gap equations, where knowledge of the ghost and gluon Euclidean dressing functions is sufficient to obtain other two-point functions. Different approaches to this issue include diagonalising for Euclidean values of $P^2$ and extrapolating to the Minkowski region, or solving gap equations in the complex plane\cite{blank}. As will be seen, extrapolation is not viable in this case and one must work with Minkowski values of the glueball momentum. Since we have chosen to use lattice data to model the gluon and ghost propagators, this implies that we must find a parameterisation of the data which reliably carries it into the region $P_E^2 <0$. Of course this is impossible without additional information. We have thus made two models for the gluon propagator that agree in the Euclidean region but yield drastically different results in the Minkowski region (the dashed and dotted lines of Fig. \ref{fig-gluon}). 

We remark that the model of Eq. \ref{eq-B-fit-1} postulates an entire gluon dressing function, which violates causality unless colour sources are strongly localised. But this is consistent with confinement, and hence entire propagators are sometimes used in hadronic model building\cite{AvS}.

The application of the formalism developed here to the pseudoscalar glueball reveals substantial simplification. Specifically, the 
antisymmetry of the indices in Eq. \ref{eq-zeromp}
  guarantees that the contact contribution to the BS equation is zero. 
Similarly the coupling to the ghost BS amplitude is zero. This is anticipated because under parity and charge conjugation the ghost Bethe-Salpeter amplitude transforms as

\be
\chi_R(k_+,k_-) = \Pi_R \, \chi_R(\tilde k_+,\tilde k_-)
\ee
and
\be
\chi_R(k_+,k_-) = {\cal C}_R \, \chi_R(k_-,k_+),
\ee
where $\Pi_R$ and ${\cal C}_R$ are the parity and charge of the state $R$. Thus spin zero states of negative parity cannot be made and even charge conjugation requires that the BS amplitude is even in $k_+\cdot k_-$ or $k\cdot P$.

Finally, although we consider the pure Yang-Mills theory in order to facilitate comparison to lattice computations of the glueball spectrum, the coupling to pseudoscalar quark-antiquark components is also suppressed. This is because the dominant term in the pion BS amplitude is known to be $E$ (see Eq. \ref{eq-pion})\cite{BSmesons}, thus
$\rlap{$\chi$}\,\chi \propto \gamma_5$. Since four Dirac matrices are required to yield a nonzero value when traced against $\gamma_5$, the coupling of the pseudoscalar glueball to the leading component of the pion is zero.
Thus the structure of the pseudoscalar glueball is dominated by the three-gluon portion of the BS equations.
The scalar glueball is more complicated: ghost and contact terms contribute and, as shown above,  there are two gluonic components in the BS amplitude.

The coupled BS equations have been solved using a number of strategies. The simplest approach assumes the `angle approximation' wherein all angular dependence of the kernel, ${\cal K}$,  is neglected. Remarkable simplification ensues and the eigenvalue problem can be easily and accurately solved. The angle approximation then serves as a useful diagnostic for the full problem. The latter has been solved by making an expansion in Chebyschev polynomials of the second kind (described in the Appendix). As an alternative, we also discretised in the $(y,k)$ plane and directly diagonalised. We found that the partial wave expansion converged quickly, requiring about the same number of waves as the order of the eigenvalue. Also, the angle approximation was very accurate for the ground state. 

The solution strategy followed that for the gap equations: the gluon and ghost lattice propagators were multiplicatively renormalised at a scale $\mu_g$ and the coupling $g(\mu_g)$ was determined by fitting to lattice gauge results. For the latter we relied on Ref. \cite{chen} who determined 

\be
M(0^{++}) = 1730 \pm 94 \ {\rm MeV} \qquad M(0^{++'}) = 2670 \pm 222\ {\rm MeV}
\ee
and
\be
M(0^{-+}) = 2590 \pm 136 \ {\rm MeV} \qquad M(0^{-+'}) = 3640 \pm 189\ {\rm MeV}.
\ee
Statistical and scale errors have been combined in quadrature. It is of note that the newer work of Ref. \cite{chen} does not quote excited state masses, so perhaps these should be taken with caution.

Given that the lattice results are not renormalised, it is tempting to implement a Wilsonian renormalisation scheme by imposing a regularisation with scale $\Lambda$ and determining $g(\Lambda)$ (and $m(\Lambda)$ if present) by fitting to observables such as glueball masses. Unfortunately we have found that the gap equations are numerically unstable when implementing this scheme, and it is necessary to render the equations finite by renormalising.

A number of surprising features emerged from the solutions. First, the effect of the ghost channel is very small and it can safely be ignored. This is perhaps related to a second peculiarity we observed, namely the ground state mass depends very weakly on the strength of the contact term, even if that strength is varied over several orders of magnitude. This strange behaviour was confirmed analytically on very small discretisation grids.


\begin{figure}[ht]
\includegraphics[width=8cm,angle=0]{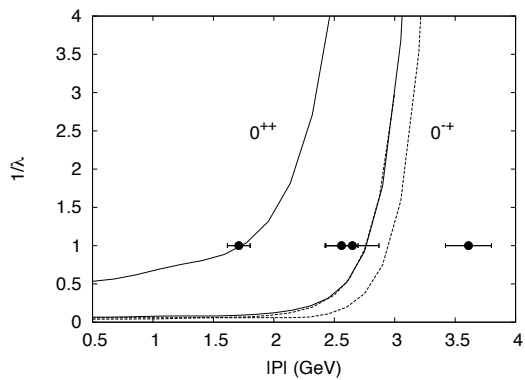}
\caption{Glueball Bethe-Salpeter Eigenvalues. Lattice data \protect\cite{chen} are represented by the horizontal bars.  Solid lines: ground states; dotted lines: first excited states. ($\mu =1$ GeV, $g(\mu) = 0.51$, intermediate vertex model).}
\label{full-gb}
\end{figure}

The sensitivity of the gluon two-point function to assumptions motivates examining a variety of three-gluon vertex models. We have found that the ground state scalar glueball mass is largely independent of the assumed form of the vertex, however, the pseudoscalar does depend on the vertex model. If one set the coupling with the scalar mass, then using the bare three-gluon vertex yields a pseudoscalar mass of 2270 MeV. Employing either the von Smekal {\it et al.} or Pennington-Wilson vertex model gives 2990 MeV. The best result is obtained if an intermediate three-gluon vertex is used where the ghost propagators are kept at their bare values in Eq. \ref{3g-vertex}. Fitting the scalar glueball mass in this case gives $g(1 \ {\rm GeV}) = 0.51$, a pseudoscalar mass of 2750 MeV, and the results shown in Fig. \ref{full-gb}. We find that the excited scalar glueball is approximately degenerate with the ground state pseudoscalar. This also held when the bare vertex model was used, and is consonant with lattice results. However, the excited pseudoscalar state is too light with respect to the lattice, indicating that the effective interaction in this channel is too soft.

All of these results have been obtained with the entire gluon propagator model.
The alternate fit (Eq. \ref{eq-B-fit-2}) does not yield physically acceptable results: the eigenvalues never cross unity for reasonable couplings and are not monotonic in the coupling strength -- indicating that the solutions are anomalous. It is thus clear that the spectrum is sensitive to the form of the analytic continuation of the gluon propagator into the Minkowski region and to the vertex models employed in the kernel.

\section{Conclusions}

We have attempted to find a consistent description of the ghost and quark propagators and the low lying scalar and pseudoscalar glueball spectrum with the Schwinger-Dyson/Bethe-Salpeter formalism and the lattice gluon propagator as input. While a reasonably good description of all of the quantities was obtained, the couplings required to obtain such descriptions were drastically different ($g(1) = 4.8$ for the quark propagator, $g(1)= 2.4$ for the ghost propagator, and $g(1) = 0.51$ for the glueball spectrum).  Furthermore, recall that gauge covariance and our renormalisation choices require that $Z_c = Z_F$. Direct computation of these renormalisation constants yields $Z_c = 1.37$ while $Z_F = 0.89$ at $\mu_g = 1$ GeV. This 35\% discrepancy again points to quite large residual truncation-dependence.
It thus appears that a consistent description of simple properties of QCD remains to be achieved. In particular, the effective strength of the gluonic interaction in the model glueball kernel is much too large. 

A variety of theoretical and practical issues bedevil this work. Amongst the former are the breaking of gauge invariance and even renormalisability that occur when truncations to Schwinger-Dyson equations are made. Finding truncation schemes that retain fundamental properties of the Lagrangian is clearly a desirable goal. On the practical side, even simple two-point functions are strongly dependent on model assumptions. As an illustration of the issues involved, the Kugo-Ojima confinement criterion specifies that the ghost dressing function $h$ approaches infinity at the origin\cite{AvS}. The lattice (and our computation) thus do not agree with this requirement, implying that something vital is missing in our understanding of the infrared properties of QCD.

There are, of course, more direct sources of possible trouble. For example, it is clearly desirable to compute the gluon propagator so that the analytic continuation to the Minkowski region can be made with some reliability. Vertex models need to be vetted and improved as much as possible. Gauge invariance needs to be maintained to the level of accuracy desired in the numerical outcomes. And the effects of topological or many-gluon terms in the interaction kernel need to be investigated. It will be of interest to include quark effects in the computation.  Lastly, an attempt at noncovariant gauges such as Coulomb or axial gauge should be illuminating\cite{ss}.

\acknowledgments
This research was supported by the U.S. Department of Energy under contract
DE-FG02-00ER41135. We thank I.L. Bogolubsky for providing lattice data and C. Aguilar, G. Eichmann, P. Tandy, C. Roberts, and D. Wilson for discussions.

\appendix*
\section{Partial Wave Decomposition of the Bethe Salpeter Equation}

The generic BS equation is

\be
\vec F(k,\hat k \cdot \hat P) = \int d^4q\, {\cal K}(q^2,k^2,P^2; x,y,w)\, \vec F(q,\hat q \cdot \hat P)
\ee
with

\be
\hat k = (\cos\theta_1, \sin\theta_1\cos\theta_2, \sin\theta_1\sin\theta_2\cos\theta_3,\sin\theta_1\sin\theta_2\sin\theta_3)
\ee
and
\be
\hat q = (\cos\phi_1, \sin\phi_1\cos\phi_2, \sin\phi_1\sin\phi_2\cos\phi_3,\sin\phi_1\sin\phi_2\sin\phi_3).
\ee
We proceed by taking $P$ in the glueball rest frame and defining $y = \cos\theta_1$, $x = \cos\phi_1$, and $w = \hat q \cdot \hat k$. It is possible (and crucial) to simplify this by choosing $\hat k$ in the 1-2 plane. Then $w$ simplifies to

\be
w = \cos\theta_1\cos\phi_1 + \sin\phi_1\sin\theta_1 \cos\phi_2 = xy + \sqrt{1-x^2}\sqrt{1-y^2}z
\ee
with $z = \cos\phi_2$. With this choice the $\phi_3$ integral is trivial and the $\theta_2$ and $\theta_3$ dependence drops out.


The partial wave decomposition is defined in terms of Chebyschev polynomials of the second type which are orthogonal over a measure appropriate for the four-dimensional integrals being considered here:

\be
\int_{-1}^{1}dx\, \sqrt{1-x^2}\, U_n(x) U_m(x) = \frac{\pi}{2} \delta_{nm}.
\ee
We define
\be
\vec F(k,y) \equiv \sum_n U_n(y) \vec F_n(k).
\ee
With this decomposition one obtains

\be
\vec F_n(k) = 4 \sum_m \int q^3 dq\, \vec F_m(q) {\cal K}_{nm}(q^2,k^2,P^2)
\ee
with
\be
{\cal K}_{nm} = \int dx \, dy \, dz\, \sqrt{1-x^2}\sqrt{1-y^2}\, U_n(y) U_m(x)\, {\cal K}(q^2,k^2,P^2;x,y,w).
\ee


\end{document}